# New exotic beams from the SPIRAL 1 upgrade

P. Delahaye[a,1], M. Dubois[a], L. Maunoury[a], J. Angot[b], O. Bajeat[a], B. Blank[c], J. C. Cam[d], P. Chauveau[a,e], R. Frigot[a], B. Jacquot[a], P. Jardin[a], P. Lecomte[a], S. Hormigos[a], O. Kamalou[a], V. Kuchi[a], B. Osmond[a], B. M. Retailleau[a], A. Savalle[a], T. Stora[f], V. Toivanen[a], J. C. Thomas[a], E. Traykov[g], P. Ujic[a], R. Vondrasek[h]

[a]*GANIL, bd Henri Becquerel, 14000 Caen, France*

[b]*LPSC, avenue des martyrs, 38026 Grenoble, France*

[c]*CENBG, 19 chemin du Solarium, 33175 Gradignan, France*

[d]*LPC Caen, Bd Maréchal Juin, 14000 Caen, France*

[e]*CSNSM, bat. 101, Domaine de l'Université de Paris Sud, 91400 Orsay*

[f]*ISOLDE, CERN, route de Meyrin, 1211 Geneva, Switzerland*

[g]*IPHC, Batiment 27, 23 Rue du Loess, 67200 Strasbourg*

[h]*ANL, 9700 Cass Avenue, Lemont, IL 60439, United States*

**Abstract**

Since 2001, the SPIRAL 1 facility has been one of the pioneering facilities in ISOL techniques for reaccelerating radioactive ion beams: the fragmentation of the heavy ion beams of GANIL on graphite targets and subsequent ionization in the Nanogan ECR ion source has permitted to deliver beams of gaseous elements (He, N, O, F, Ne, Ar, Kr) to numerous experiments. Thanks to the CIME cyclotron, energies up to 20 AMeV could be obtained. In 2014, the facility was stopped to undertake a major upgrade, with the aim to extend the production capabilities of SPIRAL 1 to a number of new elements. This upgrade, which is presently under commissioning, consists in the integration of an ECR booster in the SPIRAL 1 beam line to charge breed the beam of different 1+ sources. A FEBIAD source (the so-called VADIS from ISOLDE) was chosen to be the future workhorse for producing many metallic ion beams. This source was coupled to the SPIRAL 1 graphite targets and tested on-line with different beams at GANIL. The charge breeder is an upgraded version of the Phoenix booster which was previously tested in ISOLDE. It was lately commissioned at LPSC and more recently in the SPIRAL 1 beam lines with stable beams. The upgrade additionally permits the use of other target material than graphite. In particular, the use of fragmentation targets will permit to produce higher intensities than from projectile fragmentation, and thin targets of high Z will be used for producing beams by fusion-evaporation. The performances of the aforementioned ingredients of the upgrade (targets, 1+ source and charge breeder) have been and are still being optimized in the frame of different European projects (EMILIE, ENSAR and ENSAR2). The upgraded SPIRAL 1 facility will provide soon its first new beams for physics and further beam development are undertaken to prepare for the next AGATA campaign. The results obtained during the on-line commissioning period permit to evaluate intensities for new beams from the upgraded facility.

___________
[1]Corresponding author. e-mail: pierre.delahaye@ganil.fr.

## 1. Scientific motivations

During the past decades, the SPIRAL 1 facility at GANIL has been delivering radioactive ion beams of unique intensity and purity for physics experiments. SPIRAL 1 makes use of the so-called "Isotope Separation On Line" (ISOL) technique [1]. Ionized in ion sources, ISOL beams have an optical quality which is comparable to this of the stable ion beams. In particular, the energies accessible at SPIRAL, from a few keV to 20 AMeV for the lightest nuclides accelerated by CIME, allow for a unique variety of studies using common techniques of decay spectroscopy, Coulomb excitation, fusion, transfer and direct reactions. During the past decade a number of interesting physics results were achieved at SPIRAL 1 using these techniques, leading to published highlights addressing questions that can only be handled with reaccelerated radioactive beams. Despite remarkable achievements [2] and its status of first world-class facility, SPIRAL 1 has been technically limited to the production of radioactive ion beams of gaseous elements, thus limiting the physics opportunities of the facility.

A project of upgrade was started in 2011 to complement the radioactive ion beam production capabilities of the facility towards condensable elements (see Sec. 2). The scientific interest of the upgrade of SPIRAL 1 was sustained by different calls for letter of intents, which gathered a large response from the international nuclear physics community. In total, more than 100 beams were discussed during a workshop dedicated to SPIRAL 1, 2/3rd of which are using a new Target Ion Source System (TISS), which couples the SPIRAL target with a FEBIAD ion source [3]. First experiments using the new TISS should run during 2019. The upgrade opens on the one hand numerous opportunities with reaccelerated beams with innovative setups such as AGATA, MUGAST or ACTAR. On the other, it offers rich perspectives with low energy beams for weak interaction physics and beta/rare decay studies, in which DESIR [4] will play a major role in the upcoming years.

## 2. Upgrade layout

Since 2001, SPIRAL 1 makes use of thick graphite targets, on which the high power heavy ion beams of GANIL impinge. The radioactive fragments diffuse as atoms out of the hot target and effuse via a transfer passage to the ion source. Ionized to the required charge state by the Nanogan III Electron Cyclotron Resonance (ECR) ion source, the radioactive ions are then separated and accelerated in the Cyclotron d'Ions de Moyenne Energie (CIME) [1]. Fig. 1 presents the Target – Ion Source System (TISS) that was used so far at SPIRAL 1. The ionization technique prevents the ionization of condensable elements as the transfer tube and the ECR source is a cold assembly that stops the effusion of radioactive isotopes of condensable elements from the hot target to the ion source plasma. The SPIRAL upgrade project consisted in the development of a 1+ to n+ charge breeding system, permitting the use of versatile 1+ sources for extending the range of elements available for post-acceleration to condensable elements.

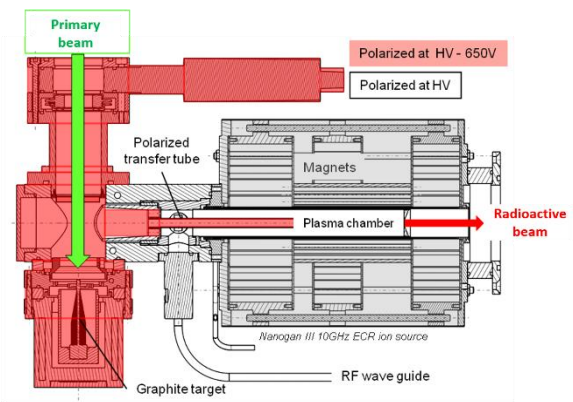

Figure 1: Target –ion source system used at SPIRAL 1 since 2001. It couples the Nanogan III ECR ion source with a graphite target via a cold transfer tube.

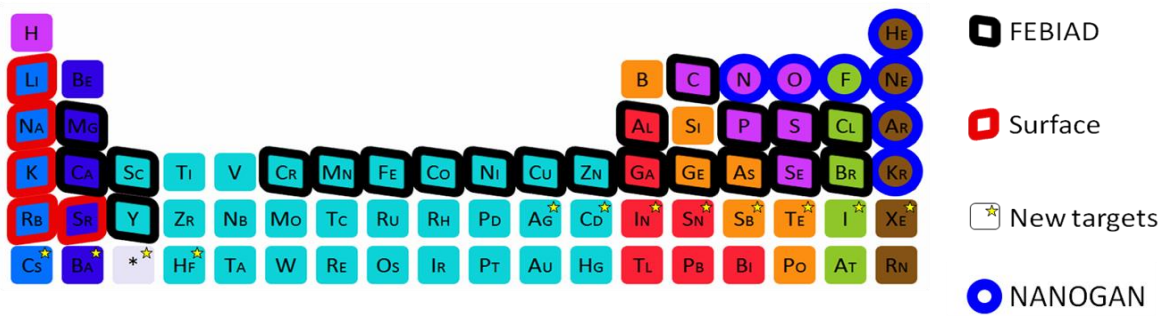

Figure 2: elements produced by the different sources envisaged after the upgrade. The FEBIAD source is a universal technique to produce 1+ beam of all elements that exhibit a melting point below 2000°C, and is therefore capable to ionize the same elements than the ECR ion sources and surface ionizers. Some of the elements that could be produced by fusion evaporation beyond Z=41 are shown by a star.

*2.1. A FEBIAD source as workhorse for the upgrade*

Compared to rare gases, radioactive atoms of condensable elements produced by the ISOL method suffer from comparatively longer sticking times on surfaces that they encounter during their transport from the target to the ion source volume. As high temperature reduces the sticking time, the condensable elements are traditionally produced using hot target and ion source assemblies [5]. Typical ion sources used in this case are hot surface sources, such as surface ionization sources, Forced Electron Beam Induced Arc Discharge sources (FEBIAD sources also called "hot plasma" sources), and Resonant Ionization Laser Ion Source (RILIS). An alternative to this scenario was tested at GANIL in the past by coupling directly ECR sources to targets at the SIRa test bench [6,7]. Due to the diversity of the beams requested in the letters of intent, a FEBIAD source recently improved at ISOLDE, the so-called VADIS [8] was found to be the ion source of condensable elements to be coupled in priority with the SPIRAL targets. The FEBIAD source can deliver a number of beams in addition to those traditionally produced by surface ionization and ECR ion sources, as can be seen on Fig. 2. It is a first step towards a universal beam production that the more sophisticated, but cleaner RILIS could complement at a later stage.

*2.2. New targets*

The upgrade was also aiming at extending the safety authorizations for a number of combinations of primary beams and targets. The previous authorizations were limited to projectile fragmentation on graphite targets. In the frame of the upgrade, these authorizations were extended to other target materials for fragmentation and fusion evaporation reactions. The following primary beam – target combinations are now authorized:

•  $^{12}$C (up to 95 MeV/A) to $^{238}$U (up to 8 MeV/A) impinging on thick graphite targets, as it previously was

•  $^{12}$C ($2.10^{13}$ pps) at 95 MeV/A impinging on thick targets of Z up to 41 (Nb)

•  $^{12}$C (95 MeV/A@$2.10^{13}$pps) to $^{238}$U (8 MeV/A@Imax.) on thin targets without restriction on Z

A Nb target and fusion evaporation targets are presently being developed [9,10] and should be tested in the coming years. The fusion evaporation target benefits from a collaboration with IPN Orsay [10].

*2.3. ECR charge breeding*

At reacceleration facilities, charge breeding is a technique consisting in the injection of a 1+ beam into the plasma of an Electron Beam Ion Source (EBIS) or Electron Cyclotron Resonance Ion Source (ECRIS) and of its subsequent multi-ionization for reaching the charge state required by the post-accelerator [11]. The charge breeding is a competitive and cost effective solution compared to the stripping foil technique. It has benefited from multiple studies within the European research and technical development programs FP5 and FP6, and from the development of facilities using this technique at ISOLDE, TRIUMF or more recently ANL. Because of the continuous mode of operation and of the intrinsic resolving power of the cyclotron CIME, an ECR charge breeder has been found better suited than a pulsed EBIS. The ECR charge breeder of the SPIRAL upgrade is a Phoenix booster donated by the Daresbury Laboratory. It has been tested at ISOLDE from 2003 to 2008 [12,13] and upgraded since to reach the best performances with this device.

## 3. New beams from the FEBIAD TISS

### 3.1. TISS development

The FEBIAD TISS couples standard SPIRAL 1 1200W C targets to the VADIS via an ohmic heated tantalum transfer tube (Fig. 3). The TISS was developed in a staged approach. The early versions were first tested online in 2011 using the SIRa test bench at modest primary beam power. They demonstrated the efficient ionization of radioactive isotopes of 8 new elements: the alkali Na, K, metallic Mg, Al, Fe, Cu, Mn and halogen Cl elements. The TISS lifetime was however limited due to various issues related to the dilatation of the transfer tube. In 2013, these issues were fixed by enabling a displacement of the ion source along the axis of the secondary ion beam extraction (Fig. 3) [14]. The modified TISS was then tested at nominal power (1200W of $^{36}$Ar at 95 AMeV) in the SPIRAL 1 beam lines in December 2013, leading to the first scientific results obtained at SPIRAL 1 with a FEBIAD source [15,16]. These on-line tests at SPIRAL 1 gave generally even better secondary beam intensities than expected from the measurements at SIRa, showing that the yields do not scale linearly with the primary beam power. The intensities shown in Fig. 4 were published in [17]. At the very end of the tests, the anode entered in short circuit with the cathode. This failure was attributed to the slow dissociation and metallization during operation of BeO insulators supporting the anode of the FEBIAD (see Fig. 3), as Be$^+$ ions were visible in high amount (1µA), dominating the ion source spectrum. In order to increase the reliability of the target ion source, a program of test was rigorously followed on the SPIRAL 1 off-line test bench with the aim to select the best insulator, which could stand operation under high temperature. The results of these tests permitted to show that on the test bench boron nitride (BN) insulators permitted a stable operation over a period of about 3 weeks. These tests were realized with an important number of heating and cooling cycles, as the ion source was switched on in the morning and off in the evening for half of the tests period. The nominal and reproducible efficiencies for stable rare gases were determined (Sec. 3.3). At this stage, the target ion source was considered as ready for operation. 2016 and 2017 were used to provision the FEBIAD target ion sources for the startup of the upgrade and diverse characterizations detailed below. In April and May 2018, the SPIRAL 1 upgraded facility was commissioned for radioactive ion beam production. A beam of 95 AMeV of $^{20}$Ne first impinged on the SPIRAL 1 target for the production of a $^{17}$F beam for the E750 experiment. Unexpectedly, the anode entered again into a shortcut with the cathode very rapidly (6-8h) after the beam was tuned on target. Again, the anode insulators were identified as the probable cause for the FEBIAD breakdown, contrasting with the observations done on the test bench without beam on target. During discussions with a panel of experts from ISOLDE and TRIUMF, the possible explanations mentioned for the different behavior of the source on-line and off-line were the different heating /thermal conditioning of the target ion source as well as the possible deposition of direct C vapors on different parts of the source, including the insulators. In order to fix these issues, a new FEBIAD ion source was conditioned with a different configuration of heat shields permitting to cool down the insulators. A helical chicane was inserted in the transfer tube to stop the direct C vapor from the irradiated target (Fig. 5). With these modifications, the FEBIAD target ion source could be run with first a low power (100-200W) $^{40}$Ca beam and finally a high power (>800W) $^{36}$Ar beam, both at 95 AMeV without failure. The secondary beam intensities measured for the latest test are shown in Fig. 6. As it clearly appeared later, the diagnostic used to center the primary beam on target was malfunctioning during the commissioning tests of 2018. The wrong steering caused lower yields than expected. As it was observed later with a Nanogan TISS delivering $^{14}$O to the E744 experiment, a factor of 50 to 100 improvement of the production yield could be obtained at the SPIRAL identification station [18] by tuning the primary beam steerers to correct for the faulty diagnostic. The wrong steering might even have been a possible cause for the early failure of the FEBIAD TISS with $^{20}$Ne.

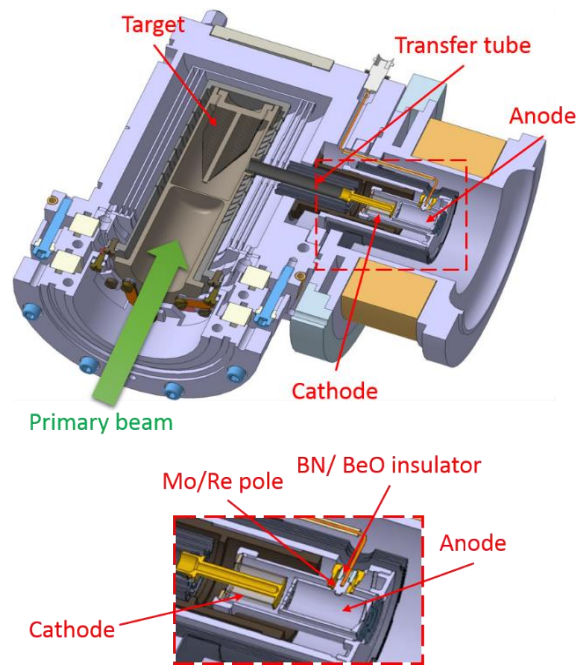

Figure 3: FEBIAD TISS. The bottom insets is a zoom on the ion source critical parts. See text for details.

The FEBIAD target ion source is being presently further consolidated for final adjustment on the test bench, before being recommissioned online in the upcoming running period. In addition to the aforementioned modifications, a few adjustments are tested. In particular the material of the poles supporting the fragile insulators, originally in Mo, are replaced by Re (see inset of Fig. 3), which has a lower thermal conductivity. These Re poles will permit to further protect the insulators from the hottest parts of the ion source. With these adjustments, it is believed that the FEBIAD target ion source, will have most chances to behave very well online.

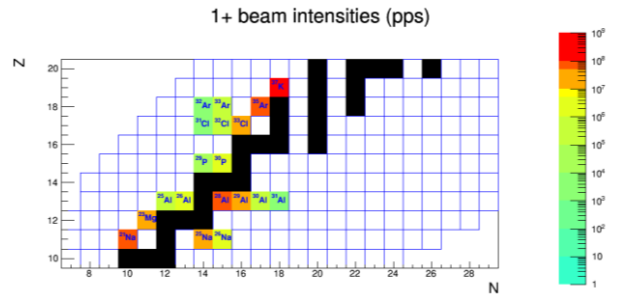

Figure 4: Secondary intensities measured with the SPIRAL 1 FEBIAD TISS with $^{36}$Ar at nominal power (~1200W) in 2013.

### 3.2. Secondary beam intensities

The 1+ beam intensities measured in 2013 and 2018 from the FEBIAD TISS at high power with a beam of $^{36}$Ar at 95 AMeV are shown in Fig. 4 and 6. The yields measured in 2018 are considered as very conservative, as it was understood later that the primary beam positioning was not controlled during the on-line commissioning, resulting in degraded production conditions.

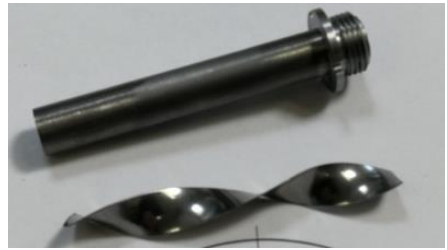

Figure 5: photograph of the transfer tube and associated helical chicane.

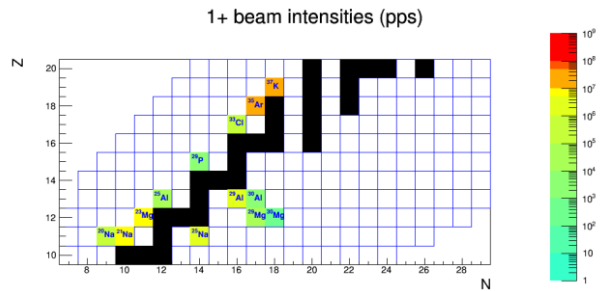

Figure 6: Secondary intensities measured with the SPIRAL 1 FEBIAD TISS with $^{36}$Ar at nominal power (800 - 1200W) in 2018. A wrong steering of the primary beam is invoked as the reason for the lower yields as compared to 2013.

An analysis of the data measured in 2013, briefly presented in [17], permitted to deduce from the measured intensities the typical release times and ionization efficiencies for the different elements observed with a Ge detector at the 1+ beam identification station. This analysis bases itself on the empirical release parametrization used in [19] for the ISOLDE yields. From the ionization, transport and release efficiencies deduced from this analysis, and using the EPAX V2 parametrization [20] for fragmentation cross sections, projections can be made for the isotopes of the different elements produced so far with the FEBIAD TISS. The results of these projections are shown in Fig. 7, making an optimized use of different primary beams available at GANIL. These projections were used as guideline for "day 1" beams, available for the first call for proposals of experiments with the SPIRAL 1 upgraded facility. For experiments using reaccelerated beams, these 1+ beam intensities have to be folded with charge breeding (~5-10%, see Sec. 4) and acceleration efficiencies (~20%), so that the resulting accelerated beam intensities are typically a factor 100 lower.

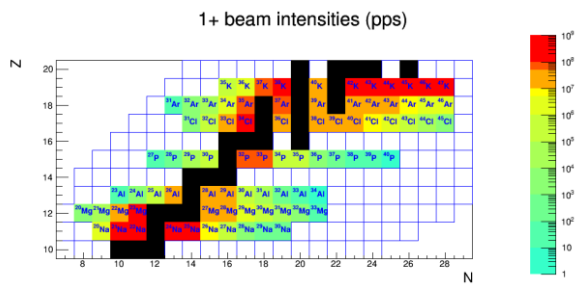

Figure 9: Projections for secondary intensities for radioactive isotopes produced by the FEBIAD TISS for the "day 1" experiments.

More speculative extrapolations have been done for the production of radioactive isotopes for other elements typically produced by FEBIAD sources, as for instance observed at ISOLDE. For these elements, the release efficiencies have been estimated from the parameterization of on-line data obtained at ISOLDE [19], PARNNE [21], and if not available from diffusion – effusion coefficients found in the literature, using the parametrization of diffusion and effusion described by Kirchner in [22]. The FEBIAD ionization efficiencies are extrapolated from measured data with noble gases. The results of such extrapolations are shown for the fragmentation of the different stable beams available at GANIL and using new target materials for fragmentation and fusion evaporation in Fig. 8 and 9 respectively. The following fragmentation targets were considered: SiC, CaO, NiO, Nb using a $^{12}$C beam at 95 AMeV and at the highest power available at GANIL (3.6 kW).The fusion-evaporation cross sections have been estimated using the procedure described in [23] for a number of target material – stable beam combinations. The plot shows only the best combination. The same release efficiencies as for the fragmentation targets have been assumed, which is a priori conservative as for this reaction mechanism many tricks can be used in order to speed up the transport of radioactive atoms to the ion source [9]. A list of secondary beam intensities as obtained using these different estimates are available on the GANIL – SPIRAL 2 website [24].

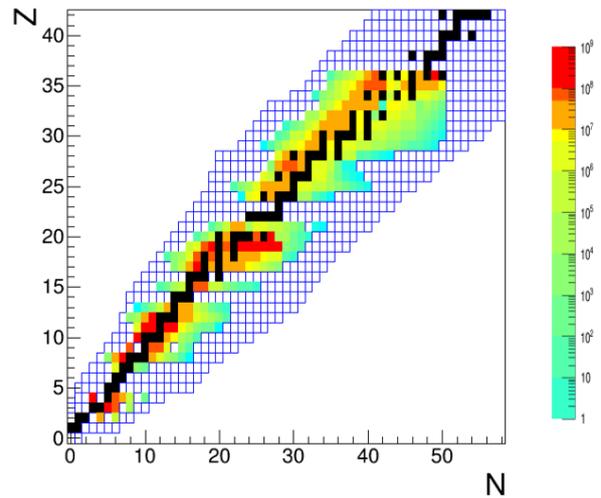

Figure 8: Extrapolation of secondary beam intensities (before post-acceleration) for the fragmentation of different stable beams available at GANIL [24].

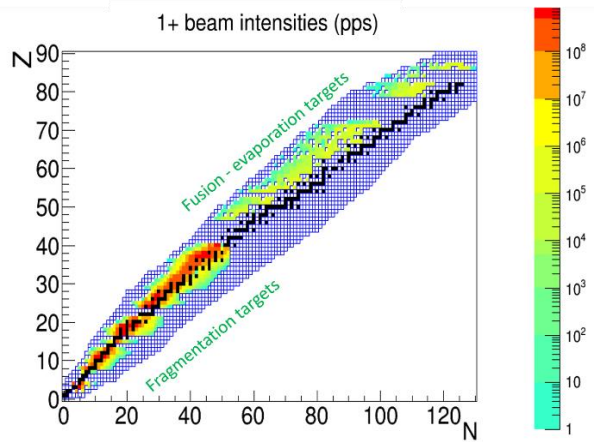

Figure 7: Extrapolation of secondary beam intensities (before post-acceleration) for the fragmentation of different targets and fusion evaporation. See text for more explanations.

*3.3. Ongoing investigations*

During the conditioning of the various versions of the TISS, the ionization efficiencies of the FEBIAD have been repeatedly monitored on the off-line test bench with stable rare gases. These ionization efficiencies are generally reproducing the efficiencies of the traditional MK5 of ISOLDE, which are a factor of ~4 lower than the VADIS ones quoted in [8]. ISOLDE equally reports that similarly lower efficiencies are regularly obtained on-line with the VADIS. Nevertheless, it is remarkable that the efficiencies quoted in [8] could be sometimes obtained during the conditioning of the TISS at the SPIRAL 1 test bench, over periods of a few hours. The conditions for stabilizing this enhanced ionization regime have not yet been found, and are being investigated. One of the latest manifestation of this particular regime happened during the injection of a tiny leak of Xe (Fig.10), whose flow rate enabled the control of the ionization efficiency, switching gradually from the standard low ionization regime to the enhanced ionization regime and vice-versa, for a couple of hours.

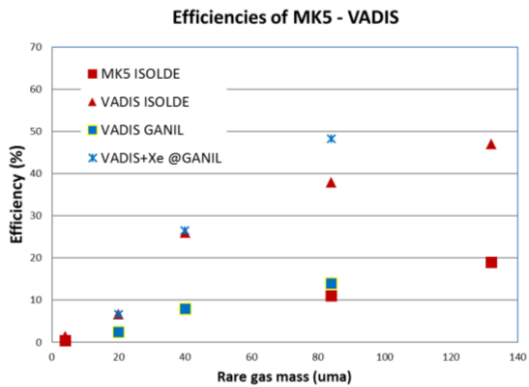

Figure 11: Ionization efficiencies measured at ISOLDE and GANIL with the different FEBIAD sources and in different ionization regimes. See text for more explanations.

The energy profile of the 1+ beam from the FEBIAD source is of high interest for the charge breeding performances. The energy acceptance of the ECR charge breeder is defined by the width of its "$\Delta V$" curve; the n+ extracted ion beam current as a function of the difference of voltage between the 1+ source and ECR charge breeder. The acceptance of ECR charge breeders is typically of 5 to 10 eV FWHM for condensable elements. The energy profiles of the FEBIAD and Nanogan III could lately be measured and compared thanks to an analyzer manufactured by LPC Caen (Fig. 11). The energy dispersion of the FEBIAD ion source is of the order of $\sigma_E \approx 1.5$ eV, which is well within the acceptance range measured with the SPIRAL 1 charge breeder [25]. It is also remarkable that the plasma potentials of the Nanogan and FEBIAD source differ considerably. The origin for the voltage shown in Fig. 11 is somewhat arbitrary, because of the lack of calibration of our test bench HV power supplies. It corresponds to the 1+ source HV voltage within 20-30 V error bar. Latest simulations show that the FEBIAD can have plasma potentials as low as -70V, with an electron beam current as high as 300 mA. This result was experimentally corroborated by the first charge breeding of a radioactive ion performed at SPIRAL 1 as described in Sec. 4.

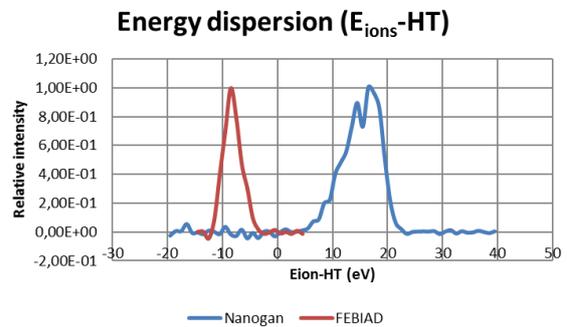

Figure 10: energy profiles of the Nanogan and FEBIAD ion sources.

In order to identify the most important mechanisms that governs the ionization performances of the FEBIAD, simulations combining the SIMION software for calculating the trajectory of charged particles (electrons and ions) in fields, and FEMM 4.2 for accounting for space charge effects, are being undertaken. At present, these simulations reproduce the plasma potential reasonably well, as the latter defines the energy profile of the 1+ beam (mean energy and dispersion) of the extracted beam. Possible ionization regimes by pure electron impact or in a plasma will have then to be investigated to attempt to reproduce the order of magnitudes of the measured efficiencies.

## 4. ECR charge breeder performances

Based on an early version of the Phoenix charge breeder used at ISOLDE for charge breeding studies [12,13], the SPIRAL 1 charge breeder has been upgraded in the frame of the EMILIE project [26] to benefit from the latest advances in the field [27,28]. The upgraded charge breeder is using ultra-high vacuum compliant components, including an Al plasma chamber. It is capable of 2 frequency heating, although only a single 14 GHz frequency heating was used so far. At injection, an iron plug has been symmetrized. The injection optics now include an electrostatic triplet and a mobile injection electrode. The puller is also mobile. The gas injection system is directly feeding the plasma chamber.

Fig. 12 presents the charge breeding efficiencies obtained on the off-line LPSC test bench [25] and in 2018 at SPIRAL 1 [29] for Na, K and Rb alkali ions, with He as support gas. The efficiencies are a factor 2 to 5 higher than those recorded at ISOLDE [12,13]. The charge breeding performances for Na were lately benefiting from the work carried out at LPSC [30]: by fine tuning the magnetic field with iron rings and the extraction electrode position, the efficiency was improved by a factor about 2-3 compared to what was previously measured at SPIRAL 1 [29].

During the radioactive ion beam commissioning in May 2018, the charge breeding of $^{37}K^{1+}$ to $^{37}K^{9+}$ was successfully demonstrated. The secondary intensities of the 1+ FEBIAD TISS beams and charge bred beams were successively transported to the identification station. An efficiency of 5.3% was obtained after a very short period of beam tuning (couple of hours) by monitoring the activity of the rather short-lived (1.225s) $^{37}K$ isotopes at the SPIRAL identification station. For such short tuning time, the tuning only concerned the 1+ beam optics and voltage difference between the ECR charge breeder and FEBIAD anode (ΔV). With an unusually high electron beam current (300 mA) emitted from the FEBIAD cathode, the optimized ΔV was of the order of 80V. Compared to usual optimized ΔV observed with surface ionization sources used for the off-line commissioning of the charge breeder (7-12V) [25,29], this value pinpoints a FEBIAD plasma potential of the order of 70V. Just prior to the injection of the radioactive ion beam, an efficiency of 10% was recorded for the charge breeding of a stable $^{40}Ar^{1+}$ beam from the FEBIAD to $^{40}Ar^{8+}$, which is a factor of ~2 below the efficiencies measured with a properly tuned charge breeder. With optimized RF power, gas flow and magnetic confinement, it is quite likely that the $^{37}K$ charge breeding efficiency would have attained a similar value as those measured with the off-line surface ionization source at LPSC and SPIRAL 1.

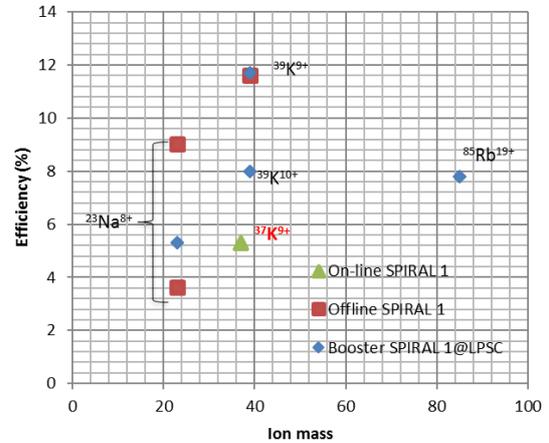

Figure 12: Charge breeder efficiencies obtained at LPSC and SPIRAL 1 with the upgraded Phoenix charge breeder.

## 5. Conclusions

The on-line commissioning of the SPIRAL 1 upgraded facility has started. As there was no charge breeding *and* subsequent acceleration of a radioactive ion beam delivered to an experiment, the full operation of the upgraded facility has not yet been fully demonstrated. Nevertheless, and despite early technical difficulties, the individual elements and combinations, which are required for the upgraded facility, are all operational. Beams have been successfully delivered from the FEBIAD TISS alone, from the FEBIAD TISS *and* charge breeder, from the charge breeder *and* CIME. The ultimate commissioning step, consisting in the successful delivery of beam to an on-line experiment, is being prepared. Final adjustments with the FEBIAD TISS are being undertaken on the off-line test bench. Charge breeding R&D is being pursued, with the aim of further improving the charge breeding efficiencies of light beams [30], and investigate the critical parameters on which the charge breeding time depends. The combination of FEBIAD TISS with an ECR charge breeder is a rather unusual one for reacceleration facilities. Experience will be gained over the years with this peculiar system. As seen with the operation of Nanogan for many years, the CIME cyclotron intrinsic resolving power permits to purify in many cases the light (A<20) beams. For heavier masses, a case – by – case study has to be done. The test of different stripping foils behind CIME to separate masses around $^{56}$Ni delivered from the FEBIAD TISS and reaccelerated by CIME to energies of 10-12 AMeV will be undertaken in the coming months. For low energy beams, the High Resolution Separator [31] and the different beam purification techniques in ion traps such as PIPERADE [32], all present at DESIR, will be essential elements for the success of the associated experimental program.

## Acknowledgments

This project has received funding from the European Union's Horizon 2020 research and innovation programme under grant agreement No 654002.

## References

[1]  A. C. C. Villari et al., Nuc. Phys. A 787 (2007) 126c-133c, and references therein.
[2]  A. Navin, F. De Oliveira Santos, P. Roussel-Chomaz and O. Sorlin, Jour. Phys. G: Nucl. Part. Phys. 38 (2011) 24004.
[3]  https://indico.in2p3.fr/event/12296/
[4]  B. Blank, B. Pramana - J Phys (2010) 75: 343.
[5]  R. Kirchner, Nucl. Instrum. and Meth. B 204 (2003) 179.
[6]  N. Lecesne, Etude de la production d'ions radioactifs multichargés en ligne, PhD Thesis, Université de Caen (1997).
[7]  P. Jardin et al., Rev. Sci. Instrum. 75 (2004)1619.
[8]  L. Penescu, R. Catherall, J. Lettry, and T. Stora , Rev. Sci. Instrum. 81(2010) 02A906.
[9]  V. Kuchi et al., to appear in these proceedings.
[10] P. Jardin et al., to appear in these proceedings.
[11] P. Delahaye, Nucl. Instrum. Meth. B 317 (2013) 389.
[12] P. Delahaye et al., Rev. Sci. Instrum.  77 (2006) 03B105 .
[13] M. Marie-Jeanne, PhD thesis, Université Joseph Fourier, 2009.
[14] O. Bajeat et al, Nucl. Instrum. Meth. Phys. B 317 (2013) 411.
[15] J. Grinyer et al., Phys. Rev. C92 (2015) 045503
[16] J. Grinyer et al., Phys. Rev. C 91 (2015) 032501(R)
[17] P. Chauveau et al, Nucl. Instrum. Meth. B 376 (2016) 35.
[18] G. F. Grinyer et al., Nucl. Instrum. Meth. A741 (2014) 18
[19] S. Lukic et al., Nucl. Instr. Meth. Phys. Res. A 565 (2006) 784.
[20] K. Sümmerer and B. Blank, Nucl. Phys. A 701 (2002) 161.
[21] B. Roussière et al., Nucl. Instr. Meth. B 246 (2006) 288.
[22] R. Kirchner, Nucl. Instrum. Meth. B 70 (186) 1992.
[23] B. Blank, G. Canchel, F. Seis and P. Delahaye, Nucl. Instrum. Meth. B, 416 (2018) 41.
[24] https://u.ganil-spiral2.eu/chartbeams/
[25] L. Maunoury et al., Rev. Scient. Instrum. 87 (2016) 02B508.
[26] P. Delahaye et al.,  Rev. Scient. Instrum. 87 (2016) 02B510.
[27] R. Vondrasek et al., Rev. Scient. Instrum. 83 (2012) 113303.
[28] P. Delahaye, L. Maunoury and R. Vondrasek, Nucl. Instrum. Meth. A 693(2012)104.
[29] L. Maunoury et al., 2018 JINST 13 C12022.
[30] J. Angot et al., Recherche et développement sur le Booster de charges Phoenix au LPSC, Journée des accélérateurs de Roscoff, 2017.
[31] T. Kurtukian Nieto et al., Nucl. Instrum. Meth. B 317 (2013) 284
[32] E. Minaya Ramirez et al., Nucl. Instrum. Meth. B 376 (2016) 298
[33] P. Ascher et al., EPJ Web of Conf. 66 (2014) 11029